\documentclass[12pt]{iopart}
\usepackage{iopams}
\usepackage{setstack}

\usepackage[english,british]{babel}
\usepackage{graphicx}
\usepackage{hyperref}
\usepackage{epstopdf}

\newcommand{\ie}{i.\,e.\ }

\newcommand{\eg}{e.\,g.\ }

\newcommand{\cf}{cf.\ }

\newcommand{\im}{\mathrm{Im}}
\newcommand{\sgn}{\mathrm{sgn}}

\begin{document}
\title[Two-colour self-ordering of polarizable particles]{Scattering approach to two-colour light forces and self-ordering of polarizable particles}
\author{S. Ostermann$^{1}$, M. Sonnleitner$^{1,2}$ and H. Ritsch$^{1}$}
\address{$^{1}$ Institute for Theoretical Physics, University of Innsbruck, Technikerstra\ss e 25, A-6020 Innsbruck, Austria}
\address{$^{2}$ Division for Biomedical Physics, Innsbruck Medical University, M\"{u}llerstra\ss e 44, A-6020 Innsbruck, Austria}
\ead{Helmut.Ritsch@uibk.ac.at}
\begin{abstract}
Collective coherent scattering of laser light by an ensemble of polarizable point particles creates long range interactions, whose properties can be tailored by choice of injected laser powers, frequencies and polarizations. We use a transfer matrix approach to study the forces induced by non-interfering fields of orthogonal polarization or different frequencies in a 1D geometry and find long range self-ordering of particles without a prescribed order. Adjusting laser frequencies and powers allows to tune inter-particle distances and provides a wide range of possible dynamical couplings not accessible in usual standing light wave geometries with prescribed order. In this work we restrict the examples to two frequencies and polarisations but the framework also allows to treat  multicolour light beams with random phases. These dynamical effects should be observable in existing experimental setups with effective 1D~geometries such as atoms or nanoparticles coupled to the field of an optical nanofibre or transversely trapped in counterpropagating Gaussian beams.
\end{abstract}

\section{Introduction}\label{intro}
Coherent interference of light scattered from different particles in an extended ensemble of polarizable point like particles leads to important modifications of the forces on the particles as well as to new inter-particle light-forces, even if the light fields are far detuned from any optical resonance~\cite{courteille2010modification,bender2010observation}. While a full 3-D treatment certainly leads to a very rich and complex dynamics~\cite{douglass2012superdiffusion}, key physical effects can already by discussed in effective 1D geometries. One particularly interesting example are atoms in or close to 1D~optical micro structures~\cite{zoubi2010hybrid,vetsch2010optical} as \eg an optical nanofibre, where even a single atom can strongly modify light propagation and forces~\cite{domokos2002quantum,horak2003giant}. In a milestone experiment Rauschenbeutel and coworkers recently managed to trap cold atoms alongside a tapered optical fibre~\cite{vetsch2010optical} and related setups predict and demonstrate strong back-action and inter-particle interaction~\cite{chang2012cavity,lee2013integrated,goban2012demonstration} leading to the formation of periodical self-ordered arrays~\cite{chang2013self,griesser2013light}. Alternatively, in free space interesting dynamical effects of collective light scattering were recently predicted and studied in standard 1D~optical lattices of sufficient optical density~\cite{deutsch1995photonic,asboth2008optomechanical}. One could also consider arrays of optical membranes to study such effects.

In this work we extend an existing model~\cite{deutsch1995photonic,asboth2008optomechanical,sonnleitner2011optical,sonnleitner2012optomechanical} towards light configurations with multiple frequencies and polarizations of the fields illuminating the particles. In particular, this includes a new class of geometries where crystalline order can be dynamically generated and sustained even without prescribing a standing wave lattice geometry. As a generic example the polarizations of two counter--propagating fields can be chosen orthogonally, such that incident and scattered fields do not directly interfere. Light scattering thus occurs for both fields independently and the forces on the particles can simply be added up. However, any structure forming by the scattering of one field component will be seen by all other fields and thus change their scattering properties and the induced forces. On the one hand this mediates nonlinear interaction between the different fields while on the other hand it generates inter particle interactions throughout the sample, inducing a wealth of nonlinear complex dynamical effects. Besides such dynamic self-ordering phenomena, we also study the possibilities to induce tailored long range interactions via multicolour illumination and collective scattering of particles trapped in prescribed optical lattice potentials.  

\begin{figure}
\centering
\includegraphics[scale=0.4]{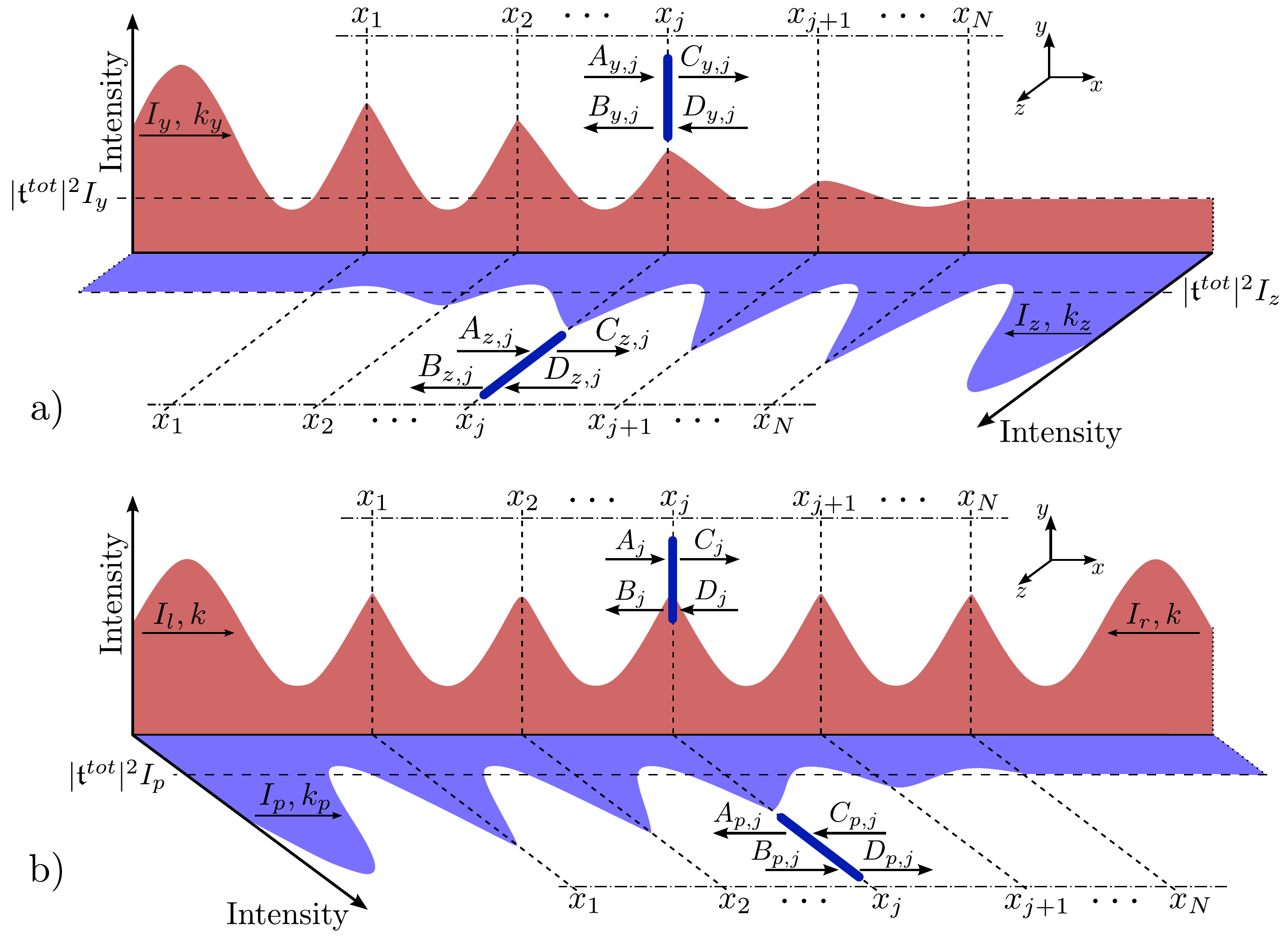}
\caption{Sketch of the intensity distribution of two light fields of orthogonal polarization and different colour propagating through a 1D array of thin beam splitters located at positions $x_1\dots x_N$. The upper graph (a) shows illumination from two sides with beams of orthogonal polarization, while the lower graph (b) shows a symmetric standing wave trap (red) perturbed by an extra field with orthogonal polarization (blue).}
\label{setup}
\end{figure}

This work is organized as follows: First we introduce the basic definitions and dynamical equations of the well--established generalized multiple scattering model for light forces~\cite{deutsch1995photonic,asboth2008optomechanical,sonnleitner2011optical,sonnleitner2012optomechanical} and extend this framework to support multiple polarizations and frequencies. This formalism is then applied to an orthogonal beam trap consisting of an array of particles modelled as beamsplitters irradiated by two counter propagating beams of orthogonal polarization and possibly different wavenumbers, \cf \fref{setup}a. For two beam splitters we analytically derive conditions on the intensity ratios and wavenumbers to trap or stabilize them at a given separation. These results are then numerically extended to higher particle numbers.

As a generalization and connection to usual experimental setups for optical lattices we then analyze how an additional beam polarized orthogonally to a prescribed standing wave perturbs the trapped particles and induces peculiar interaction patterns, \cf \fref{setup}b in 1D optical lattices.
%
\section{Multiple scattering approach to multicolour light propagation in linear polarizable media} \label{model}
It is now well established that propagation of far detuned light through a one dimensional atomic lattice or an array of dielectric membranes can be well described in a plane wave approximation with multiple scattering by a corresponding series of beam splitters~\cite{deutsch1995photonic,asboth2008optomechanical,sonnleitner2011optical,sonnleitner2012optomechanical}. A very analogous situation arises when the light is transversely strongly confined by optical structures so that scattering dominantly occurs along a preferred direction.
   
The spatial dynamics of the electric field $E(x,t)=E(x)\exp(-i \omega t)$ is then described by the 1D Helmholtz--equation
\begin{equation}
	(\partial_x^2+k^2)E(x)=-2 k \zeta E(x) \sum_{j=1}^N \delta(x-x_j),
	\label{helmholtz}
\end{equation}
where $N$ denotes the total number of beam splitters at positions $x_1,\dots,x_N$; $\zeta:=k\eta \alpha/(2\varepsilon_0)$ is a dimensionless parameter proportional to the atomic polarizability $\alpha$, the wavenumber $k=\omega/c$ and the density of particles combined to a single beam splitter, $\eta$. The plane wave solution between two beam splitters, $x\in (x_j,x_{j+1})$, then reads

\begin{equation}
	E(x) = C_j e^{i k (x-x_j)} + D_j e^{-i k (x-x_j)} \equiv A_{j+1} e^{i k (x-x_{j+1})} + B_{j+1} e^{-i k (x-x_{j+1)}},
	\label{planewave}
\end{equation}
the amplitudes $A_j$, $B_j$ left and $C_j$, $D_j$ right of the beam splitter at position $x_j$ (\cf~\fref{setup}) are related by the linear transformation matrix $M_{BS}$, with
\begin{equation}
\left( \begin{array}{cc}
C_j\\
D_j
\end{array} \right)
=
\left( \begin{array}{cc}
1+i\zeta & i\zeta\\
-i\zeta & 1-i\zeta
\end{array} \right)
\left( \begin{array}{cc}
A_j\\
B_j
\end{array} \right)
=:M_{BS} \cdot
\left( \begin{array}{cc}
A_j\\
B_j
\end{array} \right).
\label{TM}
\end{equation}

From~\eref{planewave} we read off the propagation matrix
\begin{equation}
\fl
\left( \begin{array}{cc}
A_{j+1}\\
B_{j+1}
\end{array} \right)
=
\left( \begin{array}{cc}
e^{i k (x_{j+1}-x_j)} & 0\\
0 & e^{-i k (x_{j+1}-x_j)}\\
\end{array} \right)
\left( \begin{array}{cc}
C_j\\
D_j
\end{array} \right)
=:
M_p(x_{j+1}-x_j) \cdot
\left( \begin{array}{cc}
C_j\\
D_j
\end{array} \right)
\label{prop}
\end{equation}
The values of the electric fields then are fixed by the incoming beam amplitudes $A_1$ and $D_N$. The total reflection and transmission coefficients are calculated from the total transfer matrix of the setup and give the remaining amplitudes at the boundaries as $B_1 = \mathfrak{r}_1^{tot} A_1 + \mathfrak{t}^{tot} D_N$ and $C_N = \mathfrak{r}_2^{tot} A_1 + \mathfrak{t}^{tot} D_N$ self--consistently~\cite{sonnleitner2012optomechanical}. Using Maxwell's stress tensor~\cite{jackson1999classical} yields the time averaged force per unit area on the j$^{th}$ beam splitter as~\cite{asboth2008optomechanical}
\begin{equation}
\mathcal{F}_j=\frac{\epsilon_0}{2}\left(|A_j|^2+|B_j|^2-|C_j|^2-|D_j|^2\right).
\label{force}
\end{equation}

This simple but powerful formalism to calculate the fields and forces on single atoms, atom clouds or other dielectric media such as membranes or elastic dielectrics allows to describe complex dynamics such as self--organization or even laser cooling in any effective 1D geometry~\cite{xuereb2009scattering,Chang2012ultrahigh,ni2012enhancement,xuereb2012strong}.\\

Previous approaches were limited to a single frequency and polarization in a counter propagating geometry. Here we show that it is straightforward to generalize the beam splitter method to allow for multiple frequencies and polarizations. The field propagating in the $x$-direction shall then be written as
\begin{equation}
\mathbf{E}(x,t) = E_y(x) \exp(-i \omega_y t) \mathbf{e}_y + E_z(x) \exp(-i \omega_z t) \mathbf{e}_z,
\label{Efield}
\end{equation}
where $E_y(x)$ [$E_z(x)$] is defined as the component polarized in the direction of $\mathbf{e}_y$ [$\mathbf{e}_z$] oscillating with frequency $\omega_y=k_y c$ [$\omega_z=k_z c$]. We want to emphasize that writing the the total field as a sum of linearly polarized fields in~\eref{Efield} is an arbitrary choice. None of the upcoming conclusions would change if we chose another orthogonal basis system (\eg circular polarizations).

The main assumption of this work is that the particles do not scatter photons from one mode into the other. As long as this is fulfilled, the beam splitter model can be employed for each component independently. This assumption is obviously correct if the beam splitters are made of non-birefringent materials as they are used in many optomechanical experiments.

If the beam splitters are assumed to be single atoms one has to take additional care as these typically have tensor polarizabilities. In this case one would choose counterpropagating circular polarized waves in equation~\eref{Efield} because the two modes then address different atomic transitions. If we additionally assume sufficiently large detuning for each field, we can also neglect mixing due to spontaneous emissions into other Zeeman-levels. The polarizability then loses all spin and polarization dependencies resulting in a scalar quantity

This is why the coupling parameter $\zeta$ introduced in equation~\eref{helmholtz} is proportional to a linear atomic polarizability and the wavenumber. For a generalization of the beam splitter method to multi-level atoms with tensor polarizability we refer to a work by Xuereb et al.~\cite{xuereb2010scattering}.  In this work we will assume that the atomic polarizability $\alpha$ is the same for $k_y$ and $k_z$, hence $\zeta_z= \zeta_y k_z/k_y$. Of course, a more realistic scenario is easily possible within our framework but it would add unnecessary complexity here. Our central goal is the study of multiple scattering dynamics and not the effect of optical pumping and polarization gradients.\\

In the following chapters we will study how the introduction of different frequency fields  provides new prospects to manipulate arrays of particles, ranging from equidistant lattices to individually tuned inter-particle distances as well as the design and control of motional couplings. 
%
\section{Light forces in counter propagating beams with orthogonal polarization}
In this section we explore forces and dynamics of a 1D lattice geometry modelled by a chain of beam splitters at distances $d_j:=x_{j+1}-x_{j}$ irradiated from both sides by light with orthogonal polarizations ($\mathbf{e}_y$ and $\mathbf{e}_z$) and possibly distinct frequencies ($\omega_{y}$ and $\omega_z$), \cf~\fref{setup}a. In contrast to a standard optical lattice setup as treated before~\cite{deutsch1995photonic,asboth2008optomechanical} no a priori intensity modulation due to wave interference is present and we start with a fully translation invariant field configuration. Hence the light field itself does not prescribe any local ordering and only multiple light scattering from the particles themselves creates local trapping forces. Due to the translation invariance of the setup no stable particle configuration can be expected. However, the coupled particle field dynamics still induce relative order. Hence our central goal is to find conditions, when the light forces induced by two non--interfering beams are nevertheless sufficient to obtain stationary stable particle arrays and how this spontaneous crystal formation arises.

\subsection{Stability conditions for two beam splitters}\label{two}

To get some first insight, we start with the simplest nontrivial example of two beam splitters at a distance $d=x_2-x_1$. The intensities from the left and right beam are given as $I_y=\frac{c\varepsilon_0}{2}\,|A_{y,1}|^2$ and $I_z=\frac{c\varepsilon_0}{2}\,|D_{z,2}|^2$, respectively. Here we chose the convention that all variables with index $y$ correspond to light polarized in direction of $\mathbf{e}_y$ which is injected from the left (negative $x-$ axis) and index $z$ corresponds to  $\mathbf{e}_z$  polarized light injected from the right. The individual beam splitters are counted from left to right with integer indices, hence, for example $B_{z,2}$  is the $B$-amplitude of the light field polarized parallel to $\mathbf{e}_z$ at the second beam splitter, \cf ~\eref{planewave}.

Using ~\eref{TM} and~\eref{prop} to compute the fields for any given distance $d$,  it is straightforward to obtain the total force on each beam splitter by simply adding the forces generated by the light in each polarization, \ie $\mathcal{F}_1=\mathcal{F}_{y,1}+\mathcal{F}_{z,1}$ and $\mathcal{F}_2=\mathcal{F}_{y,2}+\mathcal{F}_{z,2}$. The individual forces $\mathcal{F}_{y,1}$, $\mathcal{F}_{z,1}$, $\mathcal{F}_{y,2}$ and $\mathcal{F}_{z,2}$ are obtained from~\eref{force}.

Despite the simple physical situation the corresponding general analytic solution already is rather unhandy. Thus we first restrict ourselves to real valued $\zeta$  neglecting absorption in the beam splitter or equivalently neglecting spontaneous emission in an atom fibre system. Assuming small values of $\zeta$ and dropping terms of $\mathcal{O}(\zeta^3)$ and higher, we then find the following approximate formulas for the force on the two particles:
\begin{eqnarray}
\label{f1}
\mathcal{F}_1=&\frac{2}{c}\left(\frac{I_y\,\zeta^2\,(4\cos^2(d k_y)-1)}{1+4\zeta^2\cos^2(d k_y)}-\frac{I_z\,(\frac{k_z}{k_y}\zeta)^2}{1+4(\frac{k_z}{k_y}\zeta)^2\cos^2(d k_z)} \label{f1}\right),\\
\label{f2}
\mathcal{F}_2=&\frac{2}{c}\left(\frac{I_y\,\zeta^2}{1+4\zeta^2\cos^2(d k_y)}-\frac{I_z\,(\frac{k_z}{k_y}\zeta)^2\,(4\cos^2(d k_z)-1)}{1+4(\frac{k_z}{k_y}\zeta)^2\cos^2(d k_z)}\label{f2}\right).
\end{eqnarray}

\begin{figure}
\centering
\includegraphics[scale=1.00]{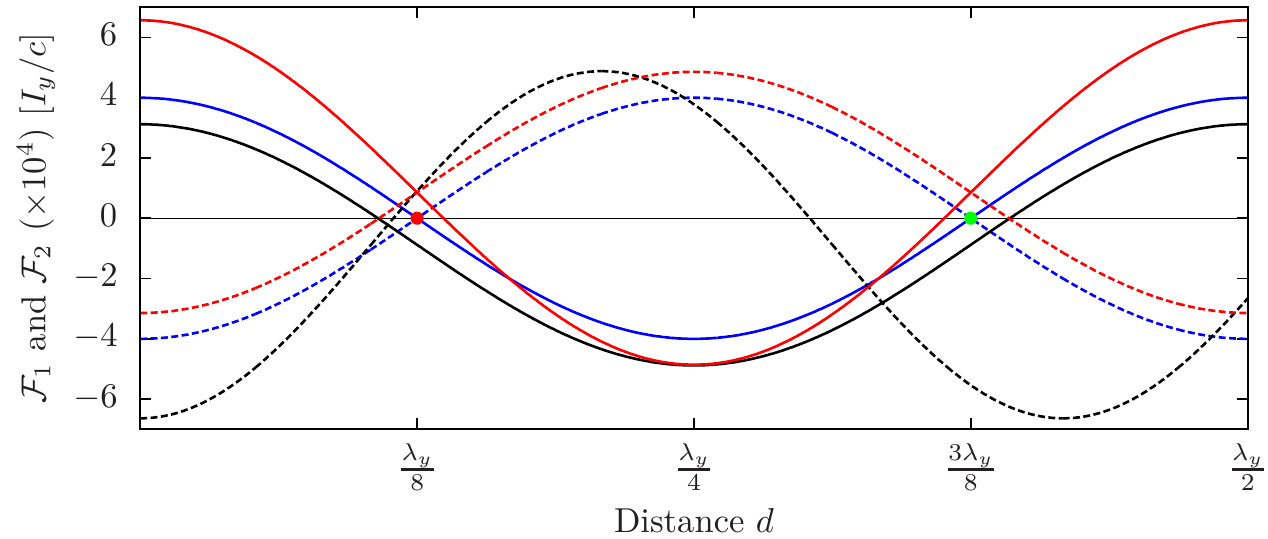}
\caption{Light force onto the  left~\eref{f1} (solid lines) and right beam splitter~\eref{f2} (dashed lines) as function of their distance $d$ for $\zeta=0.01$ and $k=k_y=k_z=2\pi/\lambda$. For equal intensity $\mathcal{P}=I_z/I_y=1$ and frequency (blue curves) the two forces add to zero and vanish at distances $d=\lambda/8$ and $d=3\lambda/8$. For asymmetric intensities $\mathcal{P}=0.7$ (red curves) we find distances with equal forces $\mathcal{F}_1=\mathcal{F}_2$ but a net center of mass force remains. The black curve shows a similar behaviour occurring for different wavenumbers $\frac{k_z}{k_y}=1.2$ of same power $\mathcal{P}=1$. The red (green) dot marks unstable (stable) stationary points. }
\label{forces}
\end{figure}

For a given set of control parameters, \ie the intesity ratio $\mathcal{P}:=I_z/I_y$ and the wavenumbers $k_y$ and $k_z$, the beam splitters will settle at a distance $d_0$ for which the two forces are equal, \ie, $\mathcal{F}_1|_{d=d_0}=\mathcal{F}_2|_{d=d_0}$, and the configuration is stable ( $\partial_d \mathcal{F}_1|_{d=d_0}>0$,  $\partial_d \mathcal{F}_2|_{d=d_0}<0$). In this case, the system can still exhibit centre of mass motion but the particles keep a constant distance. From~\eref{f1} and~\eref{f2} we see that a stable configuration in the special case of equal wavenumbers, $k=k_y=k_z$ requires
\begin{equation}
\Delta \mathcal{F}=\mathcal{F}_1-\mathcal{F}_2=\frac{4\zeta^2\cos(2 d_0 k)}{1+4\zeta^2\cos^2(d_0 k))}\frac{\left(I_y+I_z\right)}{c} = 0.
\label{selforg}
\end{equation}
Independent of the injected laser intensities, which just appear as a multiplicative factor, this corresponds to a pair distance $d^{s}_0 =\frac{(2n+1)\pi}{4 k}$ ($n \in \mathbb{N}$). Here the solutions for odd~$n$ correspond to a stable configuration, while even~$n$ leads to unstable behaviour. As numerical example we plot the full distance dependent forces for three typical sets of parameters in \fref{forces}, where stationary distances of equal force can be read of the intersection points. If these occur at zero force, the centre of mass is stationary as well. For small $\zeta$ these distances of zero force on each particle can be approximated by 
\begin{eqnarray}
\fl d_1^{\pm}=&\frac{1}{k_y}\left[\arccos \left( \pm \frac{1}{2k_y}\sqrt {\frac{k_y^2I_y+k_z^2 I_z}{I_y+(k_z/k_y)^2\zeta^2\,(I_y-I_z)}} \right)+n_1 \pi\right] \ \ \ &n_1 \in \mathbb{Z}\label{null1},\\
\fl d_2^{\pm}=&\frac{1}{k_z}\left[\arccos \left( \pm \frac{1}{2k_z}\sqrt {\frac{k_y^2 I_y+k_z^2 I_z}{I_z- \zeta^2\, (I_y-I_z)}} \right)+n_2 \pi\right] \ \ \ &n_2 \in \mathbb{Z}. \label{null2}
\end{eqnarray}
Conditions~\eref{null1} and~\eref{null2} imply $\mathcal{F}_1|_{d=d_1^{\pm}}=0$ and $\mathcal{F}_2|_{d=d_2^{\pm}}=0$, respectively.  Any solution fulfilling $d_1^{-}=d_2^{-}$ thus gives a stable and stationary configuration, where the forces on both beam splitters vanish and small perturbations induce a restoring force, as \cf \fref{forces}. In general, such solutions can only be determined numerically and are not guaranteed to exist for every set of parameters. In~\ref{int} we show that the line of argument can also be reversed and one can calculate the intensity ratios and wavenumbers needed to obtain a stable configuration for a desired distance $d$. This allows precise distance control of the particles via intensities and frequencies.

\begin{center}
\begin{figure}
\begin{minipage}{8cm}
	\centering
a)\includegraphics[scale=0.4]{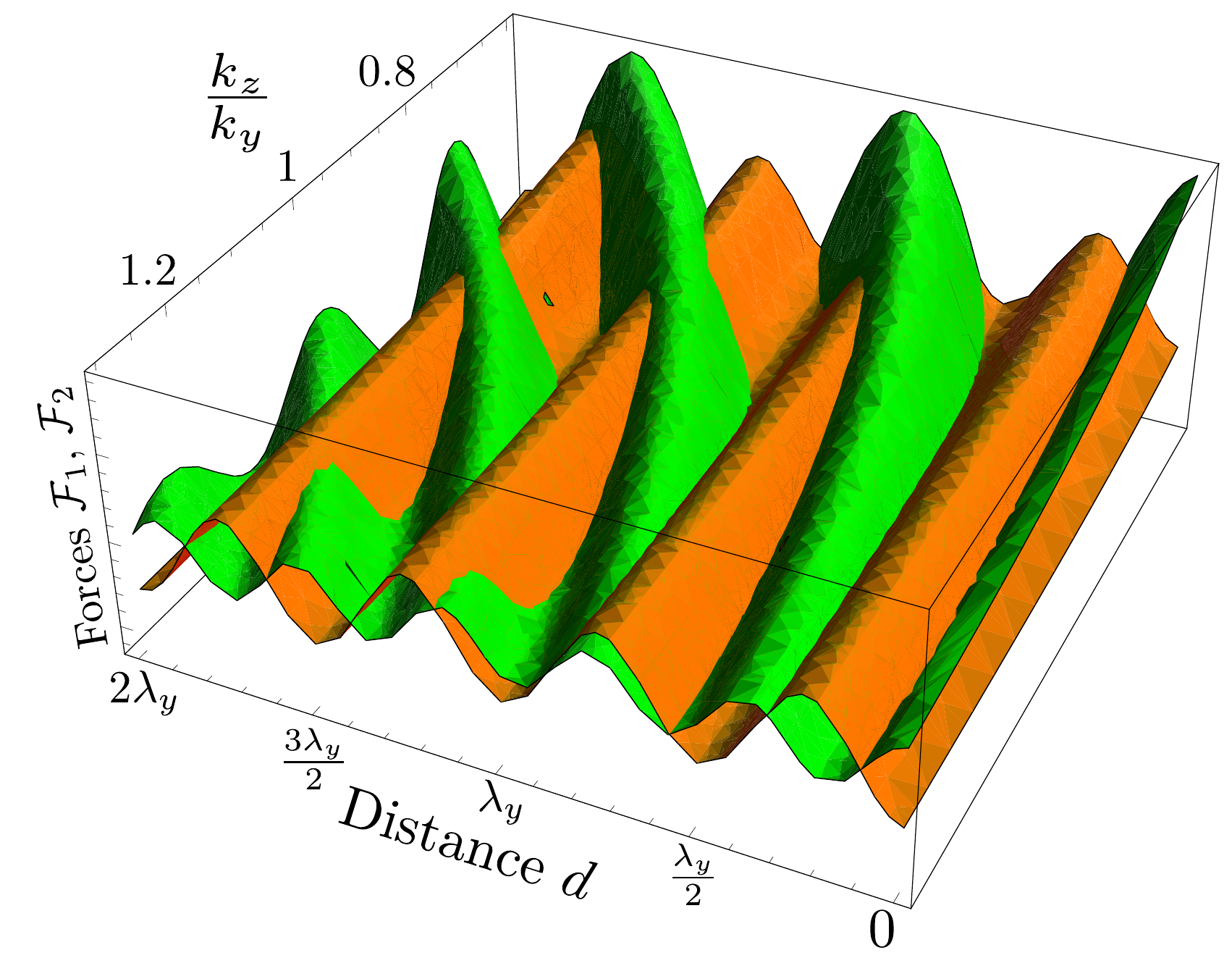}
\end{minipage}
\begin{minipage}{7cm}
	\centering
	b)\includegraphics[scale=1.2]{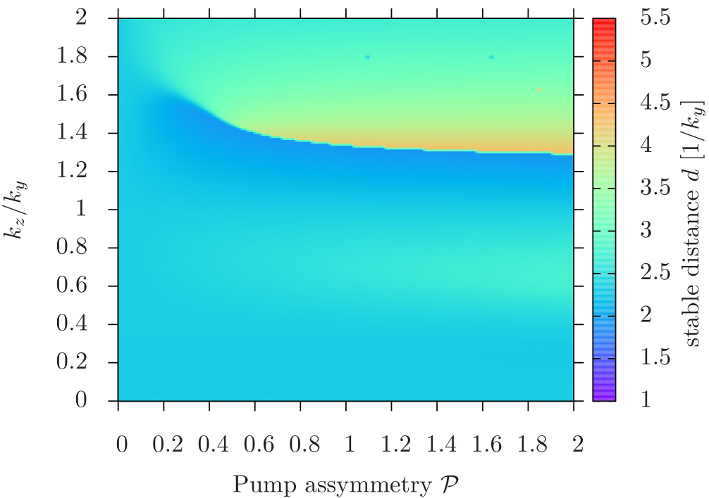}
\end{minipage}
\caption{a) Force on left (orange) and right beam splitter (green) as function of the wavenumber ratio $k_z/k_y$ and distance for two partly absorbing beam splitters with $\zeta=1/12-i/150$ and $\mathcal{P}=1$ (left figure). b) Stationary distance of two beam splitters with $\zeta=0.01$ as function of the wavenumber ratio $k_z/k_y$ and the intesity ratio $\mathcal{P}$ obtained by numerical integration of their equation of motion including an effective friction term.}
\label{2mirrors_forces}
\end{figure}
\end{center}

Let us now exhibit some more of the intrinsic complexity of the system in a numerical example. In~\fref{2mirrors_forces}a we first plot the forces on the two beam splitters as function of distance and relative wavenumber for fixed equal intensity from both sides. Clearly the intersection of the two force surfaces exhibits a complex pattern with a multitude of stationary distances which can be controlled e.g. via the chosen frequency ratio.

In an alternative approach we can numerically find a stable stationary distance of the two beam splitters as function of intensity and wavenumber ratio by time integration of their motion with some damping added, \cf\fref{2mirrors_forces}b. We see that depending on the parameters for a given initial condition the system can settle to a large range of different stationary distances, exhibiting rather abrupt jumps at certain critical parameter values. Generally a numerical evaluation requires very little effort and can be easily performed for large parameter ranges. Despite the fact that there is no externally prescribed order, the particles mostly tend to arrange at configurations with stationary distance.

\subsection{Self-ordering dynamics for higher numbers of beam splitters}
In principle, determining stationary states for a larger number of beam splitters is straightforward by first solving~\eref{TM} and~\eref{prop} for the fields and using these to calculate the forces. However, to determine a completely stationary configuration of $N$ beam splitters for a given input field configuration, we have to solve $N$ nonlinear equations to guarantee a vanishing force at each particle as function of the $N-1$ relative distances. This problem can have no or infinitely many solutions. Often one does not get an exact solution, but solutions with vanishingly small centre of mass force.  

\begin{figure}
\centering
\includegraphics[scale=0.5]{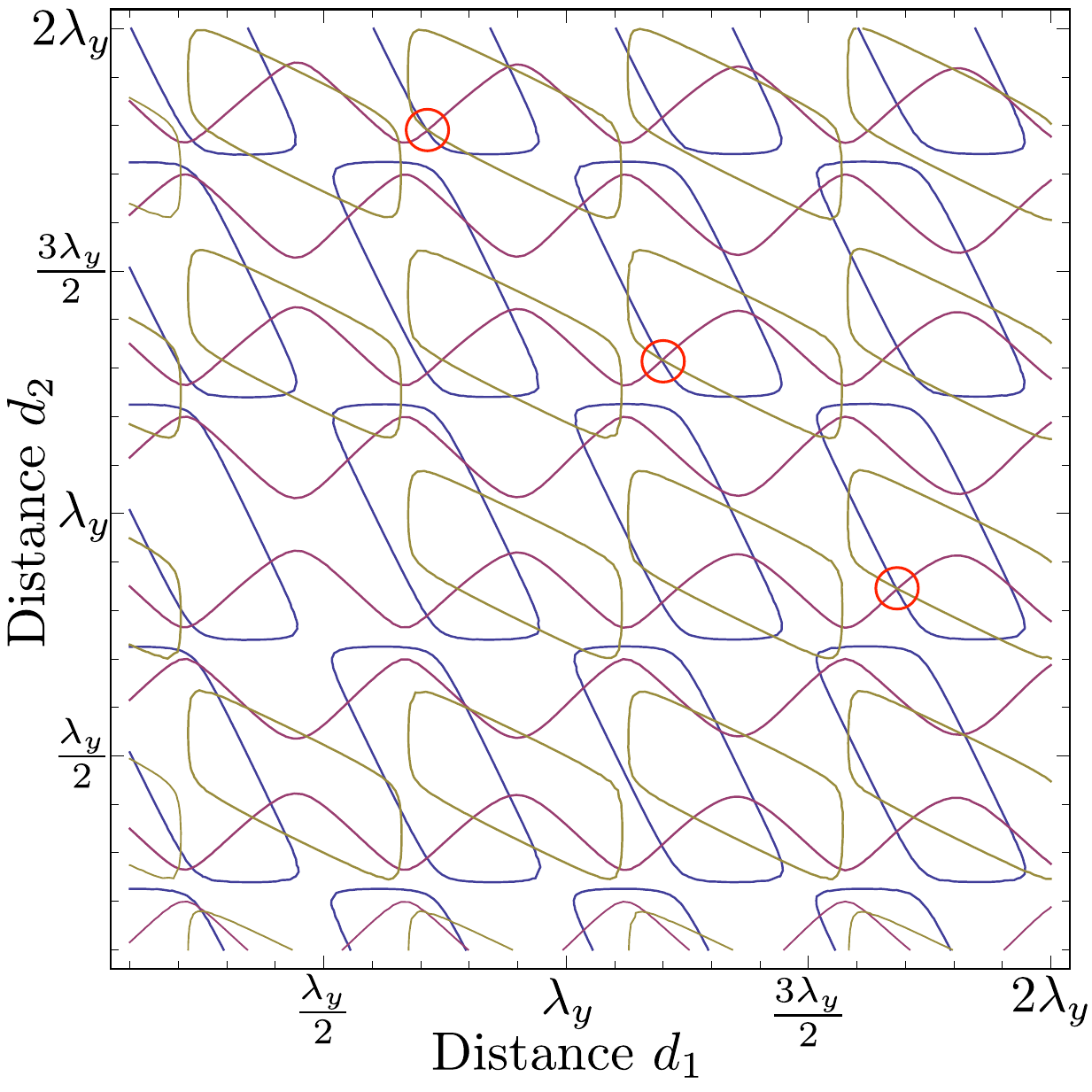}
\caption{Zero force lines for three beam splitters as function of the two distances for $k_z/k_y=1.1$, $\zeta=0.1$ and equal power $\mathcal{P}=1$. Common crossings of all three lines (red circles) denote a stationary (but possibly unstable) configuration with no centre of mass motion.}
\label{3mirrors}
\end{figure}

As a rather tractable example we plot the zero force lines as function of the two relative distances for the case of three beam splitters illuminated by light of equal power, $\mathcal{P}=1$, but different colour, $k_z/k_y=1.1$, in~\fref{3mirrors}. One finds many intersections of these lines, where two forces vanish, but only for a few distances we get triple intersections where the forces on all three particles vanish and stationary order can be achieved. These solutions then still have to be checked for stability against small perturbations to find a stable steady state.  

To investigate the dynamics of a higher number of beam splitters it is more instructive to solve the dynamical equations of motion for various initial conditions until an equilibrium configuration is reached. To arrive at a stationary solution we assign a mass to the beam splitters and add an effective friction coefficient $\mu$ in the classical Newtonian equations of motion,
\begin{equation}
m \ddot{x}_j=-\mu \dot{x}_j+F_j(x_1,...,x_N).
\label{eqm}
\end{equation}
In the following simulations we assume that the system is in the so called over damped regime, meaning that the characteristic time scale of undamped cloud motion, \ie the oscillation period, is much smaller than the relaxation time of the cloud's velocity towards a constant value due to viscous friction. Under this assumption the equations of motion~\eref{eqm} reduce to a set of differential equations of first order~\cite{asboth2008optomechanical},
\begin{equation}
\mu \dot{x}_j=F_j(x_1,...,x_N).
\label{odeqm}
\end{equation}

\begin{center}
\begin{figure}
\begin{minipage}{8cm}
	\centering
\includegraphics[scale=1.2]{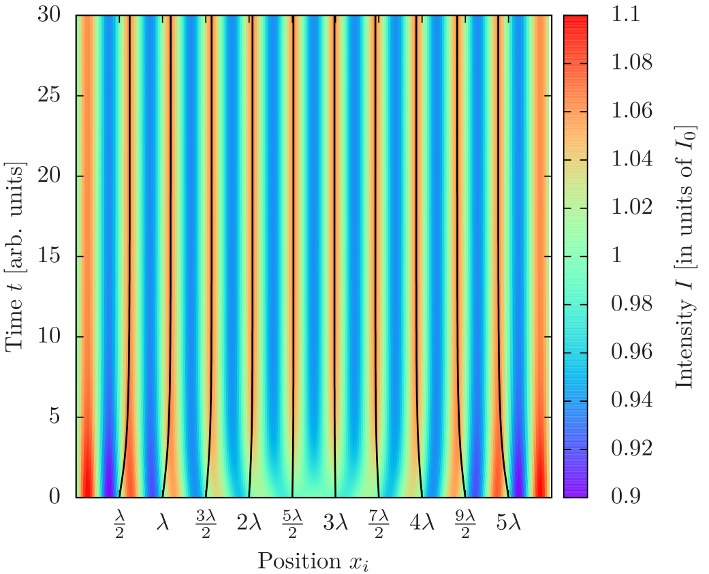}
\end{minipage}
\begin{minipage}{7cm}
	\centering
	\includegraphics[scale=0.75]{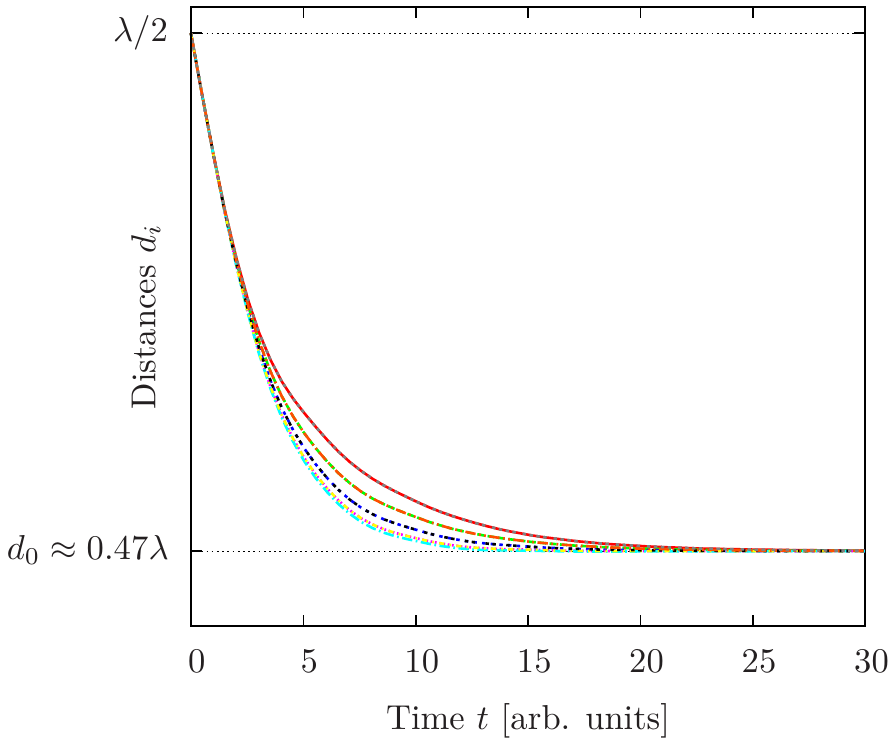}
\end{minipage}
\caption{Trajectories of $N=10$ beam splitters for $\mathcal{P}=1$, $k_y=k_z$ and $\zeta=0.01$ (left figure) started from a regular array of distance $\lambda /2$. The colour coding in the background shows the corresponding evolution of the total field intensity $I_ {tot}:=I_y+I_z$. The figure on the right depicts the change of the relative distances $d_i:=x_{i+1}-x_i$ which converge towards a stable equidistant order of reduced distance.}
\label{symtraj}
\end{figure}
\end{center}

In~\fref{symtraj} we show the solutions of~\eref{odeqm} for ten beam splitters in a simple orthogonal beam trap with $I_z=I_y$ and $k_z=k_y=k$. In a traditional standing wave trap, the beam splitters would settle at the chosen initial equidistant spacing $d_{OL}\approx\lambda/2$, \cf~\eref{latticeconstant}, which can be determined self consistently~\cite{asboth2008optomechanical}. However, for two trap beams of orthogonal polarization no prescribed periodicity is present and the particles themselves create field configurations which confine their motion through multiple scattering. Our simulations show that for a large range of operating conditions the light forces generated by two counter propagating beams with orthogonal polarizations will indeed induce an ordering of the particles, \ie multiple scattering between the beam splitters is sufficiently strong to generate a stable configuration.

\begin{figure}
\centering
\includegraphics[scale=1.00]{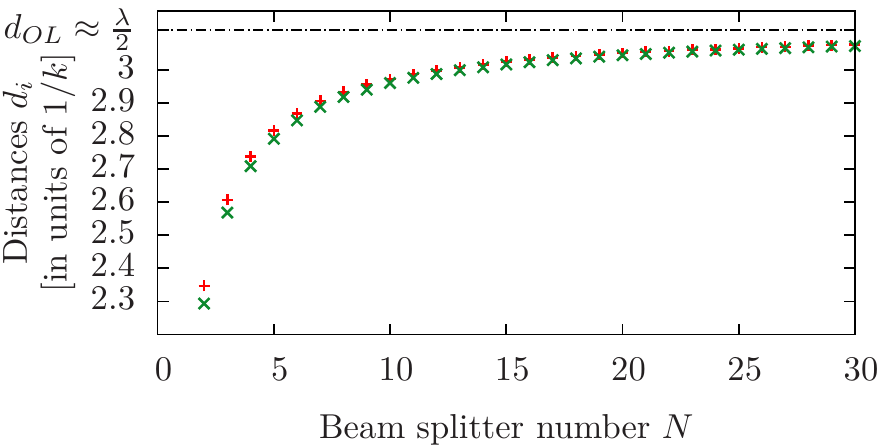}
\caption{Dependence of the relative distances $d_i$ (given in units of $1/k$) on the beam splitter number N. The parameters are chosen symmetric \ie $\mathcal{P}=1$, $k=k_y=k_z$, resulting in a equidistant lattice \cf~\fref{symtraj}. For large $N$ we observe asymptotic convergence towards the effective lattice constant as found for a standing wave configuration, \cf equation~\eref{latticeconstant} or~\cite{asboth2008optomechanical}. For the red dots we use $\zeta=0.01$. A small imaginary part of zeta (green dots) $\zeta=0.01+i 0.001$ decreases the distances but still yields stable configurations.}
\label{Ndep}
\end{figure}

Interestingly, the final distances $d_1=d_2=\dots=d_N$ converge to the same result as obtained for the standing wave optical lattice as the number of beam splitters $N$ is increased, \cf\fref{Ndep}. Thus orthogonally polarized trap beams have the same trapping properties as a standing wave setup, as $N\rightarrow \infty$. The beam splitters themselves then form an effective Bragg reflector to synthesize a standing wave configuration, which traps the particles.

\begin{center}
\begin{figure}
\begin{minipage}{7.5cm}
	\centering
\includegraphics[scale=1.2]{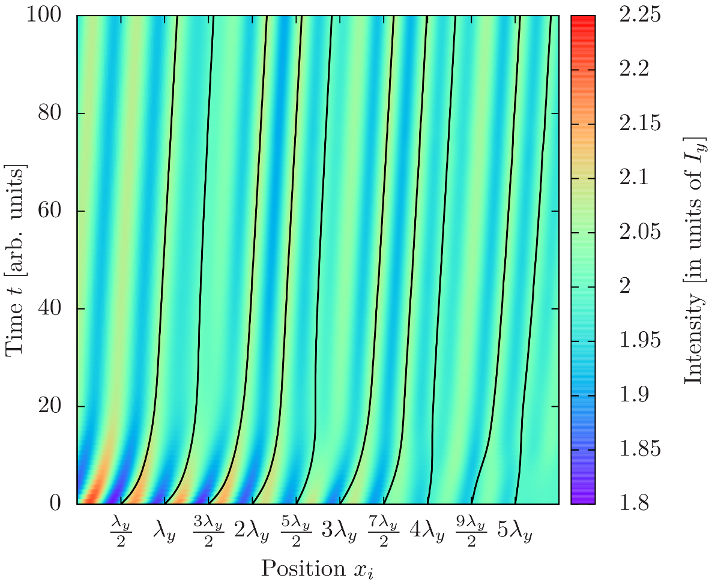}
\end{minipage}
\begin{minipage}{9cm}
	\centering
	\includegraphics[scale=1.2]{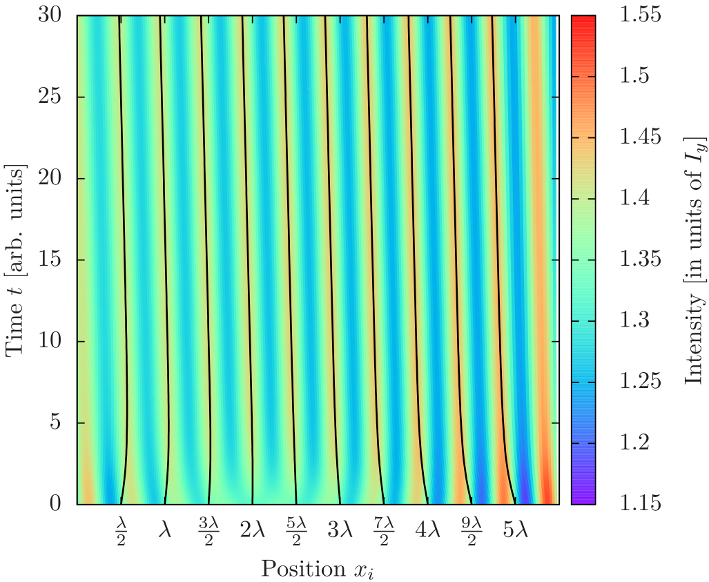}
\end{minipage}
\caption{Trajectories of $N=10$ beam splitters for $\mathcal{P}=1$, $k_z/k_y=1.3$ and $\zeta=0.01$ (left figure). The colour coding in the background shows the total field intensity $I_ {tot}:=I_y+I_z$ during the reorganization process of the system. On the right hand side the trajectories ($N=10$) for $\mathcal{P}=1.3$, $k_y=k_z$ are shown. In both cases we observe a finite centre of mass force in the long time limit, while the pattern formed is stable but is no longer equidistant.}
\label{asymtraj}
\end{figure}
\end{center}

\begin{figure}
\centering
\includegraphics[scale=1]{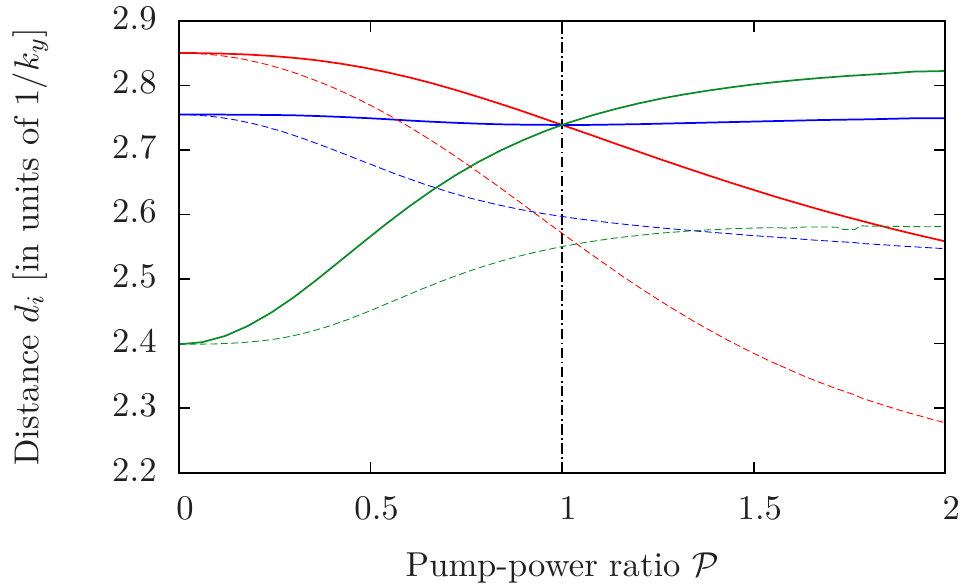}
\caption{Relative distances $d_i:=x_{i+1}-x_i$ between $N=4$ beam splitters, after the system has reorganized and stabilised, as function of intensity ratio $\mathcal{P}$. If we assume $k=k_y=k_z$ there is only one point ($P=1$) where $d_1$ (red curve), $d_2$ (blue curve) and $d_3$ (green curve) have the same values. This corresponds to a formation of an equally spaced lattice. The dashed lines show the relative distances for $k_z/k_y=1.1$, where no equidistant configuration can be realized.}
\label{dPdep}
\end{figure}

A substantially more complex behaviour is found for the case of two colour illumination with different intensities, $I_y\neq I_z$. The trajectories for some representative cases can be found in~\fref{asymtraj}. Interestingly, there is still a wide range of parameters where one obtains stationary patterns, but we generally get a non-equidistant spacing, \cf~\fref{dPdep}, and a finite centre of mass force. As above for two beam splitters, this force can be controlled via the intensity ratio to stabilize the centre of mass or induce controlled motion. Of course, the configuration not only depends on the operating conditions, but also on the initial conditions allowing for a multitude of different stationary configurations.   

In summary we conclude that the particles prefer to form crystalline structures held in place by collective multiple scattering. The more particles we have, the more complex these patterns can get and the more different solutions can exist. The complexity of the problem increases further, if one allows for a variation of the individual coupling parameters $\zeta$, \eg to represent number fluctuations of the atoms trapped at each lattice site or size variation of trapped beads. Note that although appearing similar at first sight, the mechanism is different from standard optical binding of polarizable beads, which works on transverse shaping of the incoming light with the particles acting as small lenses~\cite{dholakia2010colloquium}, which we neglected in our model. 

Let us finally note that analogous results should be obtained for a setup using two counter propagating beams of equal polarization, but sufficiently different frequencies, such that scattering between the different colours is suppressed. From the particles point of view, the interference pattern of the combined fields then oscillates so rapidly that they cannot follow and the two forces stemming from the two fields can be calculated independently. Such frequency shifts are a common method to generate 3D optical lattices by using a different frequency in each dimension. But in contrast to those cases, here we get a mutual interaction between the light intensities of the different frequency components. During the evolution the spatial shifts of the beam splitters induced by one field are seen by all other fields and influences their propagation. 
%
\section{Tailored long range interactions in a bichromatic optical lattice}\label{bichromatic}
Optical lattices for ultracold atoms are of course an extremely well established and controllable technology. In general, parameters are chosen in a way to avoid back-action of the particles on the fields. The underlying physics helps here to achieve this goal as particles tend to accumulate in zero force regions, where their influence on the lattice light is minimized~\cite{deutsch1995photonic,asboth2008optomechanical}. This is radically changed in the orthogonally polarized beam setup described above, where trapping forces are only created by the back-action of the particles on the two beams and interactions in the lattice occur via multiple collective scattering.

In the following chapter we will consider a second generic example to generate tailored long range interactions in an optical lattice. In particular we study the extra forces introduced by a second perturbation field of different wavelength in a given optical lattice formed by to two strong counter propagating beams of equal wavenumber $k$ and polarization $\mathbf{e}_y$, \cf \fref{setup}b. By adding an extra beam of different wavenumber $k_p$ and polarization $\mathbf{e}_z$ we can introduce tailored perturbations and couplings, as its gradient is generally non-zero at the positions of the original lattices sites. 

For generality we allow intesity asymmetries for the dominant standing wave field $\mathcal{P}:=I_{r}/I_{l}$ where the first indices $l$ and $r$ stand for \emph{left} and \emph{right} suggesting the direction of incidence. The intensity of the additional perturbating field is called $I_{p}$. The same index notation will also be used for the corresponding field amplitudes. 
%
\subsection{Two beam splitters in an bichromatic optical lattice}
The first relevant system to study interactions and couplings introduced by an additional field of different frequency are two beam splitters trapped at a distance $d_\mathrm{sw}$, \cf\eref{latticeconstant}, in a far detuned optical lattice at stable positions $x_1^0=x_0-d_\mathrm{sw}/2$ and $x_2^0=x_0+d_\mathrm{sw}/2$. Here $x_0$ denotes the centre of mass coordinate calculated following~\cite{sonnleitner2012optomechanical}, via
\begin{equation}
x_0=\frac{1}{2k}\left(\arccos{\left[\frac{(I_r-I_l)(1+|\mathfrak{r}|^2-|\mathfrak{t}|^2)}{2|\im(\mathfrak{r}\mathfrak{t}^*)|\sqrt{I_lI_r}}\right]}-\frac{\pi}{2} u \right)+\frac{n \pi}{k}, \ \ \ n\in\mathbb{Z}
\label{stabx0}
\end{equation}
with $u=\sgn[\im(\mathfrak{r}\mathfrak{t}^*)]$.

The reflection and transmission coefficients $\mathfrak{r}$ and $\mathfrak{t}$ of the total system derived from the total transfer matrix are
\begin{eqnarray}
\mathfrak{t}&=\frac{e^{i k (x_2-x_1)}}{\zeta ^2 \left(e^{2 i k (x_2-x_1)}-1\right)-2 i \zeta +1}\label{trans},\\
\mathfrak{r}&=-\frac{\zeta  \left((\zeta -i) e^{2 i k (x_2-x_1)}-\zeta -i\right)}{\zeta ^2 \left(e^{2 i k (x_2-x_1)}-1\right)-2 i \zeta +1}\label{refl}.
\end{eqnarray}

The incident fields of the standing wave component are assumed as $A_{l}=\sqrt{2I_l/{(\varepsilon_0c)}}\exp(ikx_1)$ and $D_{r}=\sqrt{2I_r/{(\varepsilon_0c)}}\exp(-ikx_2)$ so that the remaining amplitudes at the boundaries are $B_{l}=\mathfrak{r}A_{l}+\mathfrak{t}D_{r}$, $C_{r}=\mathfrak{t}A_{l}+\mathfrak{r}D_{r}$. This allows to calculate the lattice forces on the first and second particle, 
\begin{eqnarray}
	\fl \mathcal{F}_1&=\frac{\epsilon_0}{2}\left[|A_{l}|^2+|B_{l}|^2-|(1+i\zeta)A_{l}+i\zeta B_{l}|^2-|i\zeta A_{l}-(1-i\zeta)B_{l}|^2\right] \label{OLF1},\\
	\fl \mathcal{F}_2&=\frac{\epsilon_0}{2}\left[|(1+i\zeta)A_{l}+i\zeta B_{l}|^2+|i\zeta A_{l}-(1-i\zeta)B_{l}|^2-|C_{r}|^2-|D_{r}|^2\right]\label{OLF2}.
\end{eqnarray}

The additional perturbation field is described by the amplitudes $A_{p}=\sqrt{2I_p/{(\varepsilon_0c)}}\exp(ik_px_1)$, $B_{p}=\mathfrak{r}A_{p}$ and $C_{p}=\mathfrak{t}A_{p}$ and generates the additional forces ($\zeta_p=k/k_p\zeta$)
\begin{eqnarray}
\fl \mathcal{F}_{1p}&=\frac{\epsilon_0}{2}\left[|A_{p}|^2+|B_{p}|^2-|(1+i\zeta_p)A_{p}+i\zeta_p B_{p}|^2-|i\zeta_p A_{p}-(1-i\zeta_p)B_{p}|^2\right]\label{OLF1p},\\
\fl \mathcal{F}_{2p}&=\frac{\epsilon_0}{2}\left[|(1+i\zeta_p)A_{p,z}+i\zeta_p B_{p}|^2+|i\zeta_p A_{p}-(1-i\zeta_p)B_{p}|^2-|C_{p}|^2\right]\label{OLF2p}.
\end{eqnarray}
Here we restrict the corresponding added dynamics of the two beam splitters to small, time dependent perturbations $\Delta x_1(t), \Delta x_2(t)<<d_\mathrm{sw}$ from the equilibrium positions $x_0$ given in~\eref{stabx0}. Using $x_1(t)=x_0-d_\mathrm{sw}/2+\Delta x_1(t)$ and $x_2(t)=x_0+d_\mathrm{sw}/2+\Delta x_2(t)$ and linearising the forces for small $\Delta x_1(t)$ and $\Delta x_2(t)$ yields to the following coupled equations of motion
\begin{eqnarray}
m\Delta \ddot{x}_1(t)=-K\Delta x_1+\kappa_1 (\Delta x_2(t)-\Delta x_1(t))+F_{ext},\nonumber \\
m\Delta \ddot{x}_2(t)=-K\Delta x_2-\kappa_2 (\Delta x_2(t)-\Delta x_1(t))+F_{ext}. \label{dynsys}
\end{eqnarray}

A detailed calculation of the coefficients $K$, $\kappa_1$, $\kappa_2$ and $F_{ext}$ is shown in~\ref{lin}. The equations above correspond to two coupled harmonic oscillators driven by an external force $F_{ext}$.

The solution of the system~\eref{dynsys} can be calculated analytically yielding
\begin{equation}
\fl \left(\begin{array}{c}
	\Delta x_1(t)\\
	\Delta x_2(t)\\
\end{array}\right)
=
\left(\begin{array}{c}
	1\\
	1\\
\end{array}\right)
\left(a_1 \cos(\omega_1 t+\varphi_1)+\frac{F_{ext}}{K}\right)+a_2
\left(\begin{array}{c}
	-\frac{\kappa_1}{\kappa_2}\\
	1\\
\end{array}\right)
\cos(\omega_2 t+\varphi_2)
\label{modes}
\end{equation}
with $\omega_1=\sqrt{\frac{K}{m}}$ and $\omega_2=\sqrt{\frac{K+\kappa_1+\kappa_2}{m}}$.\\

\begin{figure}
\centering
\includegraphics[scale=1.00]{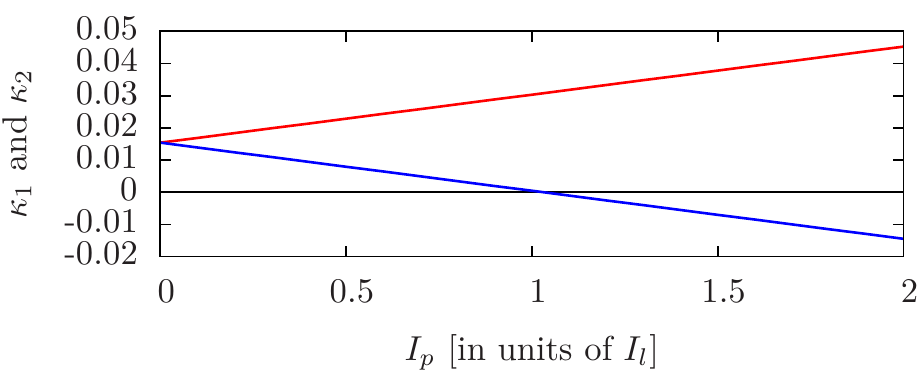}
\caption{Dependence of the coupling constants $\kappa_1$ (red) and $\kappa_2$ (blue) on the perturbation field intensity $I_{p}$ for $\zeta=\zeta_p=0.1$, $k=k_p$ and $\mathcal{P}=1$. As soon as the perturbation field is switched on, the constants differ.}
\label{coupling}
\end{figure}

Note that the coupling constants $\kappa_1$ and $\kappa_2$ here are not necessarily equal, \cf also \fref{coupling}, as there is no energy conservation enforced for the motion of the two beam splitters. Since the parameters can be chosen in a way so that the coupling constant $\kappa_1$ is equal to zero, one-sided couplings can be achieved. This means that only the motion of beam splitter number two is coupled to beam splitter number one, which does not couple to the rest of the system. The direction of this effect is governed by the direction of incidence of the perturbation beam. Besides, $\kappa_2>0$ holds for all values of $I_p$, meaning that no antisymmetric modes can be obtained if $\kappa_1<1$,~\cf\eref{modes}. Generally we find that tuning the perturbation field intensity offers a variety of different dynamics not accesible with traditional standing wave setups. This motivates a more detailed treatment of this system, using numerical methods.

\subsection{Long range coupling of beam splitters in an optical lattice}
As shown in~\cite{asboth2008optomechanical} the effective self-consistent lattice constant in a standing wave with asymmetry $\mathcal{A}:=\frac{I_l-I_r}{\sqrt{I_lI_r}}$, with $I_l:=\frac{\epsilon_0}{c}|A_{l}|^2$, $I_r:=\frac{\epsilon_0}{c}|D_{r}|^2$ adjusts to
\begin{equation}
d_\mathrm{sw}=\frac{\lambda}{2}\left(1-\frac{1}{\pi} \arccos\left[\frac{-\zeta^2\sqrt{4+\mathcal{A}^2}+\sqrt{4-\zeta^2\mathcal{A}^2}}{2(1+\zeta^2)}\right]\right).
\label{latticeconstant}
\end{equation}

\begin{figure}
\centering
\includegraphics[scale=1.00]{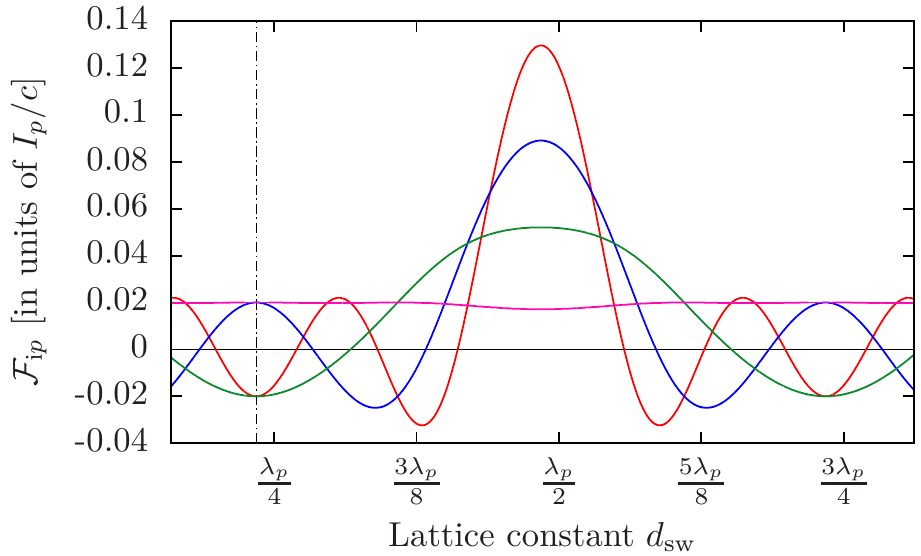}
\caption{Perturbation induced force $\mathcal{F}_{ip}$ on $N=4$ beam splitters at their unperturbed equilibrium positions in an optical lattice as function of the lattice constant $d$ for $\zeta=0.1$. The red line corresponds to the force $\mathcal{F}_{1p}$, the blue line to $\mathcal{F}_{2p}$, the green line to $\mathcal{F}_{3p}$ and the magenta line to $\mathcal{F}_{4p}$.}
\label{force_ddep}
\end{figure}

In an optical lattice, multiple scattering induces long range interactions between the particles in the form of collective oscillation modes. In the selfconsistent configuration the particles arrange at intensity maxima at minimal field gradients, so that this interaction is strongly suppressed for small perturbations.  Adding, however,  a second, perturbative field by a single running wave beam of wavenumber $k_p$ injected from one side induces an additional force on each particle perturbing the regular periodic order. This perturbation then acts back on the original standing wave field. Note that a single plane wave by it self would only add a constant force, but this force is modified by multiple scattering depending on the particle distances. As an instructive example we show these perturbing force acting on $N=4$ beam splitters  in~\fref{force_ddep}. We see that the additional force is different for all the particles and changes as a function of the lattice constant relative to the wavelength of the perturbing light. Hence, by a proper choice of parameters almost any combination of magnitudes and signs of forces on the different beam splitters can be achieved.

This can be exploited for different purposes to control and study lattice dynamics. As a first and direct application it is possible to tailor a specific field to induce oscillations of selected particles in the optical lattice by deflecting them from their equilibrium position. As shown in~\fref{force_ddep}, the force on the individual beam splitters depends strongly on the prescribed lattice constant. This means that there is a wide range of realizable dynamics as long as the lattice constant can be tuned, \cf~\eref{latticeconstant}. This can be potentially refined by simultaneous use of several perturbation frequencies.

\begin{figure}
\centering
\includegraphics[scale=0.8]{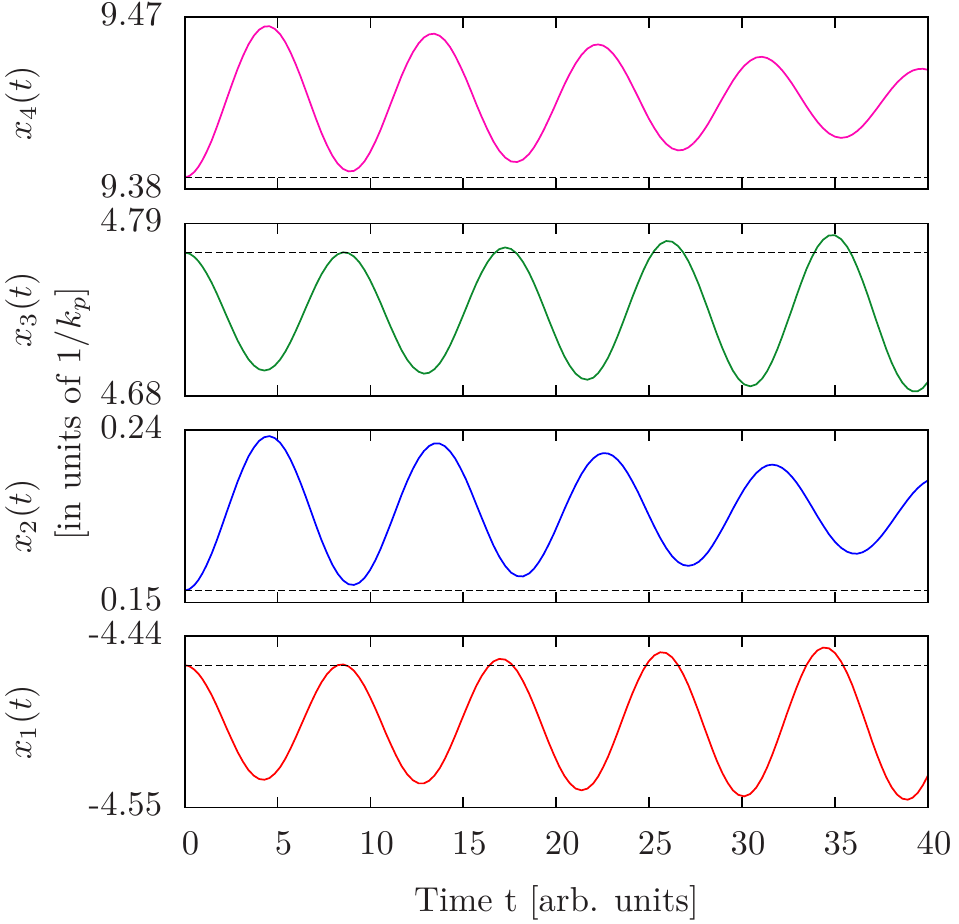}
\caption{Correlated oszillatory motion of $N=4$ beam splitters in a lattice induced by an additional perturbation field with intensity $I_{p}=I_{l}=I_{r}$, wavenumber $k_p=k$, $\zeta=\zeta_p =0.1$ and damping parameter $\mu=0.01$. The black, dashed lines show the initial  unperturbed trapping positions $x_i=n_i d_0-x_0$ ($n_i \in \{1,2,\dots,4 \}$) for $d_0=0.23\lambda_p$.}
\label{traj_pOL}
\end{figure}

For example, using the parameters from~\fref{force_ddep} we anticipate interesting behaviour for a lattice with spacing $d_\mathrm{sw}\approx0.23\lambda_p+m\pi$, $m\in \mathbb{N}$ as in that case $\mathcal{F}_{1p} = \mathcal{F}_{3p} = - \mathcal{F}_{2p} = - \mathcal{F}_{4p}$ (\cf dash-dotted line in~\fref{force_ddep}). This behaviour is verified by calculating the trajectories via~\eref{eqm}, the results are shown in~\fref{traj_pOL}. Obviously it is possible to correlate the motion of distant beam splitters in an optical lattice via the additional beam. After switching on the perturbation at $t=0$, particles number one and three show amplified oscillations, while the other's oscillations are damped.

\begin{figure}[htp]
\centering
\includegraphics[scale=0.8]{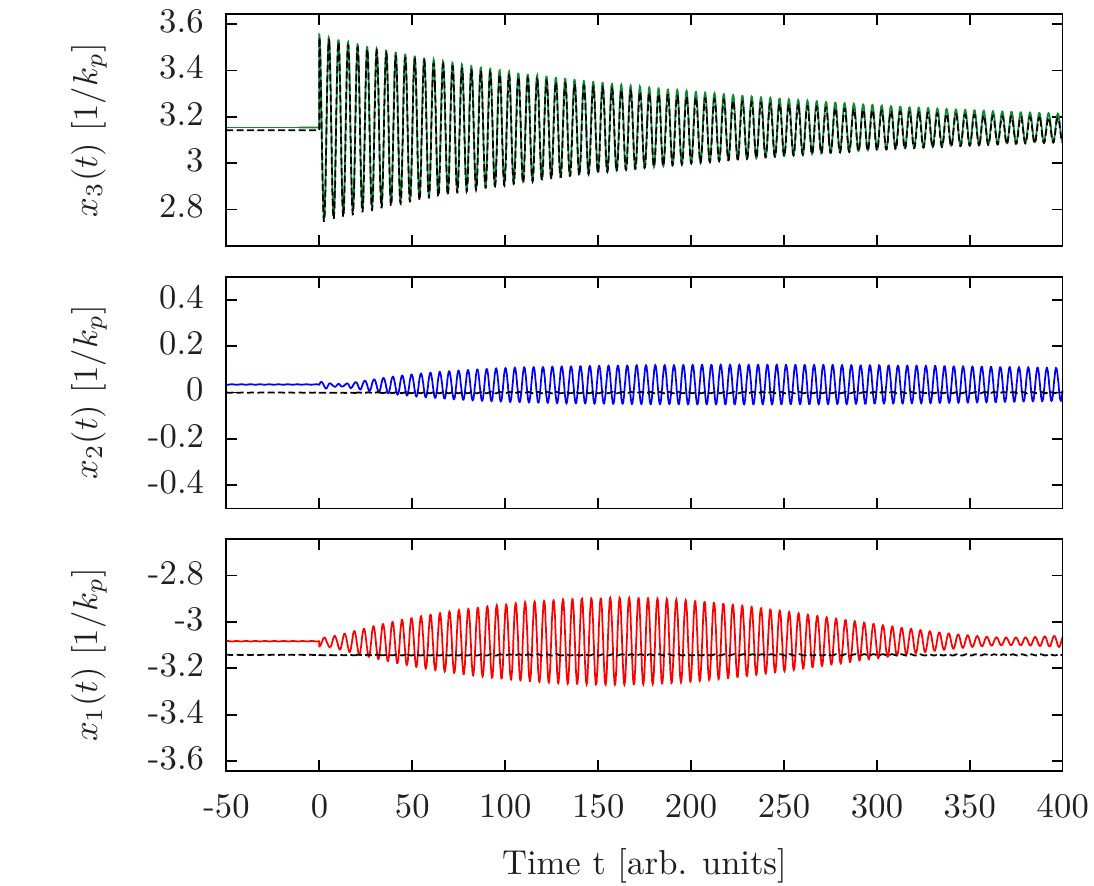}
\caption{Example plot for the resonant coupling of three beam splitters, trapped in a standing wave configuration with $d_0=\lambda_p/2$. We used $\zeta=0.01$, $\zeta_p=0.1$, $I_l=I_r=20 I_p$ and $k/k_p=0.99$. The rightmost beam splitter is displaced from the equilibrium position at $t=0$, resulting in a damped oszillation (damping parameter $\mu=0.01$). The black curves show the resulting dynamics for $I_p=0$. The green ($x_3(t)$), blue ($x_2(t)$) and red ($x_1(t)$) curves show the dynamics if the coupling field is switched on. Note the resonant coupling between $x_1$ and $x_3$.}
\label{rescoupling}
\end{figure}

In a second approach the additional field is designed to enhance interactions between selected distant areas in the lattice. As shown in~\fref{rescoupling}, exciting an oscillation of one particle weakly coupled to the standing wave field will usually have little effect on the other trapped particles. But after adding a perturbative field with carefully chosen parameters, this oscillation can be transferred to the other particles, forcing them to move along. Note that due to the fact that the additional perturbation (coupling) field is imposed from only one side, this coupling effect is not symmetric and excitations can flow in a desired direction. For example, a perturbing field entering from the left hand side will maximally transfer the motion of the rightmost beam splitter. This setup allows one to correlate the motion of particles and can even be used as a channel to transfer \eg quantum information through the lattice, very much like phonons in ion crystals.

%
\section{Conclusions}
We have shown that even in the case of non-interfering counter propagating light fields of different polarization and frequency, stable lattice configurations of particles held in space by multiple coherent scattering are possible. In contrast to conventional optical lattices the light here plays a decisive dynamical role as multiple scattering is essential to form and stabilize the structure. Compared to prescribed optical lattices the physics is much closer to the case of solids, where lattice dynamics in form of phonons not only keeps the atoms in place but also mediates long range interactions. Interestingly in conventional optical lattices such interactions can be tailored by adding additional coupling fields of suitable frequency and polarization. 
While we have performed our calculations only for 1D geometries, where a semi-analytic scattering approach can be used, similar effects should be present in 3D geometries as well. 

In general for very far detuned optical fields these effects will be rather small but their importance will grow with the size of the lattice as well as in transversally confined fields. Particularly strong effects can be expected in fields guided by nano optical devices such as nano fibres or hollow core fibres. Here even for a few particles strong interactions can be expected.

In this work we have restricted ourselves to the bichromatic case for sake of simplicity. Nevertheless one can expect even more complex dynamics for an increasing number of input fields as the forces show a more complex distance dependence. Note that here we have ignored any internal optical resonances of the particles. Working close to such resonances certainly should strongly increase the effects but also will complicate the analysis.

Let us finally mention here that the system not necessarily requires a fixed set of beam splitters as a starting point. As an alternative we can consider each beam splitter to be formed by a small sub ensemble of atoms in a 1D beam configuration, as it has been proposed before~\cite{deutsch1995photonic,asboth2008optomechanical}. In our case of noninterfering counterpropagaing beams, one can expect that under suitable conditions the cold atoms arrange in small groups forming at local field intensity maxima~\cite{griesser2013light}. Groups of atoms at certain spatial sites then commonly form beam splitters shaping a self consistent lattice structure.

In contrast to conventional lattices, backaction of the particles onto the fields is an essential part of the dynamics and the field thus strongly mediates collective interactions. Light scattering on one end of the lattice influences the lattice depth at the other end, which opens a completely new branch of ultracold atom optical lattice physics. Note that also atoms trapped in optical resonator fields~\cite{Ritsch2013Cold} exhibit similar dynamical coupling effects but in that case the backaction is strongly restricted by the resonator geometry limiting the available interaction wavenumbers.
\ack
We thank A.~Rauschenbeutel, M.~Aspelmeyer, N.~Kiesel and H.~J.~Kimble for stimulating discussions. We  acknowledge support via the Austrian Science Fund grant SFB F40 and ERC Advanced Grant (catchIT, 247024). 
%
\appendix
\section{Distance control for two particles}\label{int}
In section~\ref{two} we saw that distances with vanishing force on both beam splitters, leading to a stable or trapped configuration for given intensity ratios and wavenumbers for the special case $k=k_y=k_z$ require equal intensity from left and right. If we reverse the line of argument and ask for intensity ratios and wavenumbers where the two beam splitters can be trapped at a given distance $d$, the result allows us to study the full system, \ie different polarizations \emph{and} frequencies. In this case precise distance control is possible.

First we need to find the zeros of~\eref{f1} and~\eref{f2} with respect to $\mathcal{P}$.
\begin{eqnarray}
\mathcal{P}_1=&\frac{(4 \cos^2(dk_y)-1)(k_y^2+4k_z^2\, \zeta^2\cos^2(dk_z))}{k_y^2\,(1+4\zeta^2\cos^2(dk_y))} \label{P1}\\
\mathcal{P}_2=&\frac{k_y^2+4k_z^2\,\zeta^2\cos^2(dk_z)}{k_y^2\,(1+4\zeta^2\cos^2(dk_y))(4\cos^2(dk_z)-1)} \label{P2}
\end{eqnarray}

\begin{figure}
\centering
\includegraphics[scale=1.00]{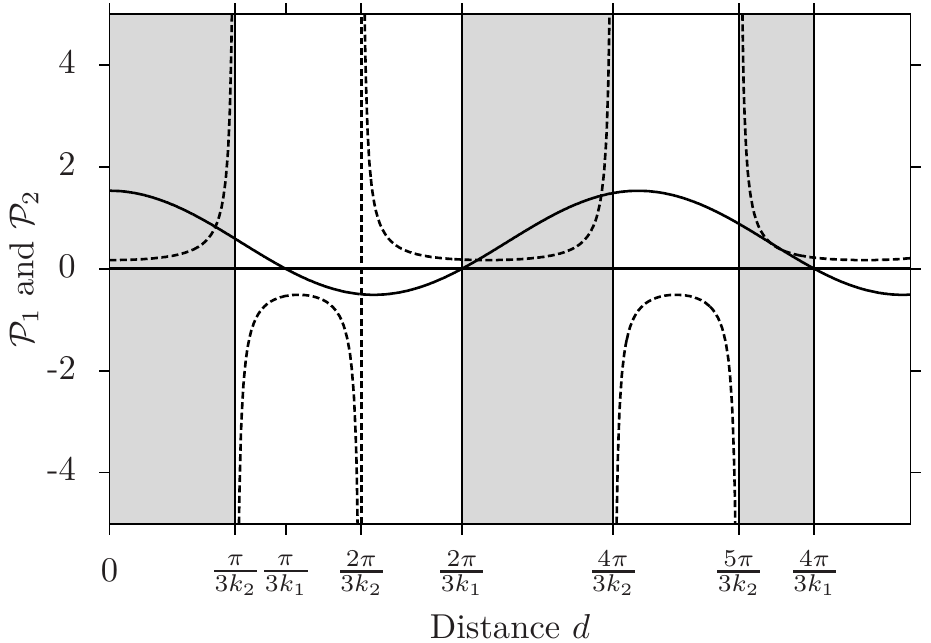}
\caption{Intesity ratio leading to stable solutions~\eref{P1} (solid line) and~\eref{P2} (dashed line) for a prescribed distance $d$ for $\zeta=0.01$ and $k_z/k_y=1.4$. The grey regions mark the physically allowed regions $\mathcal{P}_1>0$ and $\mathcal{P}_2>0$. It impossible to find intensity configurations so that particles can be trapped at distances outside of these regions.}
\label{pratio}
\end{figure}

Both solutions $\mathcal{P}_1$ and $\mathcal{P}_2$ have to be positive, which is not valid for all values of $d$ (\cf~\fref{pratio}).

To obtain a trapping condition for the wavenumbers, \ie wavenumbers where the total force $\mathcal{F}_1+\mathcal{F}_2$ vanishes, we solve $\mathcal{P}_1=\mathcal{P}_2$, finding
\begin{equation}
k_z^{\pm}=\frac{1}{d}\arccos\left(\pm \frac{\sqrt{\cos(2dk_y)}}{\sqrt{2(1+2\cos(2dk_y)}} \right)+2\pi n, \ \ \ \ n \in \mathbb{Z}.
\label{stabint}
\end{equation}
 
\eref{stabint} envolves us to calculate the needed wavenumbers to trap the beam splitters at a given distance $d$. The associated intensity ratio can be calculated via~\eref{P1} or~\eref{P2}.

Obviously there exists a wide range of parameters which allow stable and trapped configurations of beam splitters in multicolour light beams with orthogonal polarizations (or sufficiently different wavenumbers).

%
\section{Linearisation of the forces on two beam splitters in a bichromatic optical lattice}\label{lin}
In section~\ref{two} we calculate the equations of motion for two beam splitters in a standing wave geometry perturbed by an additional field with orthogonal polarization. For that purpose we use linearized forces~\eref{OLF1},~\eref{OLF2},~\eref{OLF1p} and~\eref{OLF2p}. Here we want to show how this linearization is done and how the constants $K$, $\kappa_1$, $\kappa_2$ and $F_{ext}$ can be calculated.

The force $\mathcal{F}_1$ depends only on the positions $x_1(t)$ and $x_2(t)$ of the two beam splitters. Replacing these variables via $x_1(t)=x_0-d_\mathrm{sw}/2+\Delta x_1(t)$ and $x_2(t)=x_0+d_\mathrm{sw}/2+\Delta x_2(t)$ results in a force dependent on $\Delta x_1(t)$ and $\Delta(t):=\Delta x_2(t)-\Delta x_1(t)$. Assuming small $\Delta x_1(t)$ and $\Delta(t)$ we perform a 2D Taylor approximation to first order resulting in
\begin{equation}
\mathcal{F}_1=a+b\Delta x_1+c(\Delta x_2(t)-\Delta x_1(t))
\end{equation}
where we defined real constants $a$,$b$ and $c$ which are lengthy expressions depending on the system's parameters.

The same method works for the remaining forces $\mathcal{F}_2$, $\mathcal{F}_{1p}$ and $\mathcal{F}_{2p}$, where the latter two only depend on the relative distance $\Delta(t)$.
\begin{eqnarray}
\mathcal{F}_2&=u+v\Delta x_2+w(\Delta x_2(t)-\Delta x_1(t))\\
\mathcal{F}_{1p}&=K_{1p}+K_{2p}(\Delta x_2(t)-\Delta x_1(t))\\
\mathcal{F}_{2p}&=K_{3p}+K_{4p}(\Delta x_2(t)-\Delta x_1(t))
\end{eqnarray}

Performing these tedious calculations we find that some of the obtained constants are zero ($a=u=0$) while others have the same values. We define $-K:=b=v$, $\kappa_1:=K_{2p}+c$, $-\kappa_2:=K_{4p}+w$ with $c=w$ and $F_{ext}:=K_{1p}=K_{3p}$. With this results result we get the forces used in the equations of motion~\eref{dynsys}.
%
%

\end{document}